# Molecular Orientation-Induced Second Harmonic Generation: deciphering different contributions apart

Amit Beer[1,2], Ran Damari[1,2], Yun Chen[1] and Sharly Fleischer[1,2]

[1]Raymond and Beverly Sackler Faculty of Exact Sciences, School of Chemistry, Tel Aviv University 6997801, Israel.
[2]Tel-Aviv University center for Light-Matter-Interaction, Tel Aviv 6997801, Israel.

**Abstract:**
We demonstrate and explore an all-optical technique for direct monitoring the orientation dynamics of gas phase molecular ensembles. The technique termed 'MOISH' utilizes the transiently lifted inversion symmetry of polar gas media and provides a sensitive and spatially localized probing of second harmonic generation signal that is directly correlated with the orientation of the gas. Our experimental results reveal selective electronic and nuclear dynamical contributions to the overall nonlinear optical signal and decipher them apart using the "reporter gas" approach. 'MOISH' provides new, crucial means for exploring controlled rotational dynamics via concerted terahertz and optical field excitation.

**Introduction:**
Angular control of gas molecules is a long-standing goal of physics and chemistry, aimed to lift the inherent isotropy of the gas for extracting spectroscopic signatures from the molecular-frame. Vast research efforts have successfully yielded a plethora of new observations and possibilities ranging from basic light-matter phenomena through novel spectroscopic methods for studying rotational dynamics to practical coherent control schemes and many more [1–6].
Anisotropic angular distributions are categorized as 'aligned' or 'oriented', referring to preferable distribution of the *intramolecular axis* along a specific lab-frame axis or the *molecular dipoles* toward a specific lab-frame direction respectively [7,8]. Correspondingly, alignment retains the inversion symmetry of the medium whereas orientation entails its inversion asymmetry upon orientation of the molecular dipoles toward the +z OR –z direction ('up' or 'down') [9,10]. The lifted inversion symmetry provides access to nonlinear optical responses of even orders in the field ($\propto E^{2n}$) that are otherwise forbidden in unordered gas samples. Orientation may be induced by two-color laser field [11–14], mixed field (dc+optical) [15,16] or terahertz (THz) field excitations [8,17–20] that interact resonantly via the permanent molecular dipole. We note that chiral molecules were recently shown to orient by non-resonant near-IR (NIR) pulses with twisted polarization [21,22].

THz fields induce molecular orientation by dipole-interaction with polar molecules $\hat{V} = -\vec{\mu} \cdot \vec{E}$ to create a rotational wave-packet $\psi(t) = \sum c_{J,m} \cdot e^{-\frac{i\hat{H}t}{\hbar}} |J,m\rangle$ where $\hat{H} = \frac{\hat{L}^2}{2I}$ with $\hat{L}$- the angular momentum operator, $|J,m\rangle$ are the spherical harmonic functions and their expansion coefficients - $c_{J,m}$. The rotational wavepacket $\psi(t)$ periodically reproduces itself at integer multiples of the "rotational revival time" [23,24], given by $T_{rev} = \frac{1}{2B}$ ($B$ is the rotational constant of the molecule). As the gas molecules periodically orient, they form a transient macroscopic dipole that manifests as emission of THz bursts, usually referred to as free-induction signals (or FID) [8,18,19,25] detectable via time-resolved electro-optic sampling (EOS) [26,27]. While EOS is useful for rotational spectroscopy, it provides an indirect signature of orientation since the FID follows the time-derivative of the orientation d$\langle\langle cos\theta\rangle\rangle$/dt [8,28]. Moreover, EOS practically lacks any spatial resolution since the detected FID is emitted from the entire interaction region of the THz field with the gas. The latter poses severe difficulties when multiple excitation fields are used simultaneously, e.g., NIR and THz pulses with drastically different volumetric foot-prints in the gas cell. Motivated by the need for spatially localized, direct detection of molecular orientation, we set to monitor the transient inversion asymmetry of the gas upon orientation via the SHG ($\lambda_s = 400$nm) of a NIR probe ($\lambda_{probe} = 800$nm).

The experimental approach presented and explored hereafter is closely related to the THz field-induced SHG (TFISH) method – a technique used for detection of broad-band THz fields [29–32]. TFISH relies on the nonlinear mixing of three input fields – a THz field ($E_{THz}$) and two NIR fields ($E_\omega$, $E_\omega$) via the 3rd order susceptibility, $\chi^{(3)}$, to yield a signal field at frequency $\omega_{TFISH} \cong 2\omega$. While TFISH is restricted to non-polar gasses [33] and typically performed in ambient air, *the 'MOISH' technique explored hereafter is aimed for exploring polar gas samples that are resonantly excited by THz fields* *and manifest transient orientation dynamics long after the THz field is over. Special efforts are made to decipher the electronic (TFISH) and nuclear orientation (MOISH) contributions apart at fundamental time periods in the rotational evolution of polar gasses.*

**Experimental**

The setup used in this work is similar to that reported in [32]. Very briefly, an intense THz and NIR probe beams are routed to propagate collinearly and focus inside a static gas cell equipped with a designated <1mm aperture. The latter effectively restricts the interaction length of the two beams and eases the phase-mismatch ($\Delta k$) of the generated SH signal. The NIR pulse (100fs duration, 6µJ pulse energy) is focused by a lens (f = 150 mm) such that its intensity remains well below the laser-induced plasma regime [32,34,35]. Complementary EOS measurements were performed in our home-built time-domain THz spectrometer [18,19].

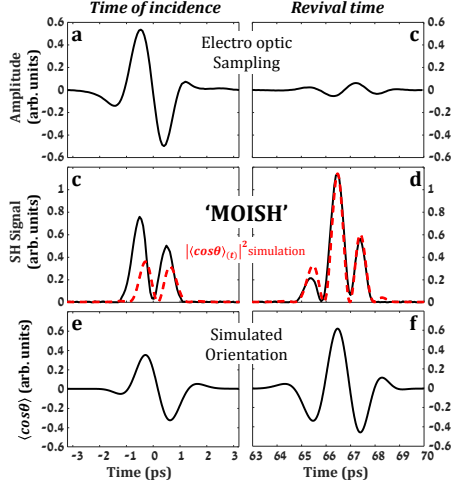

**Figure 1:** THz-induced orientation of 10torr Methyl-Iodide (CH$_3$I) at room temperature. (a) Electro-optic sampling signal. (b) Second harmonic (MOISH) signal (probe intensity 1.1·10$^{14}$W/cm$^2$). (c) Simulated orientation dynamics.

Figure 1 compares the experimental results obtained with EOS and MOISH from methyl-iodide (CH$_3$I) gas (10torr, 300K) following irradiation by a single-cycle THz field generated by optical rectification in a LiNbO$_3$ crystal [36]. In EOS, the THz field propagates through the static gas cell located at the first focus of a 4-f setup [18,19]. The THz (and succeeding FID) are re-collimated and focused onto the EO detection crystal (GaP) and sampled by a weak NIR probe. Figs.1a,b show the EOS signal with the input THz field (Fig.1a, at t=0) and the FID emission at the first revival of the gas (Fig.1b, T$_{rev}$~66ps) [28,37–39]. Figs.1c,d depict the time-resolved MOISH signal at the same respective intervals. Here, the THz field and the NIR pulse co-propagate to focus at the center of the static gas cell and the generated SH signal is recorded as a function of their delay apart. While EOS detects the THz radiation, MOISH is primarily sensitive to the degree of molecular orientation $\langle \cos\theta \rangle$. This is evident from the difference in signals' amplitudes of the incident THz field (t=0) and the FID emission at the revival time; In EOS, the incident THz field reaches the peak value of 0.5 while the FID emitted at t=T$_{rev}$ remains well below 0.1 (in the arb. units shared by Fig.1a,b). In contrast, the MOISH signal obtained at T$_{rev}$~66ps (Fig.1d) is ~50% larger than that of the incident THz field (Fig.1c), in good agreement with the simulated orientation dynamics $\langle cos\theta \rangle_{(t)}$ (Figs.1e,f). Note that the dashed red line in Figs.1c,d is the absolute value squared $|\langle cos\theta \rangle_{(t)}|^2$ of the simulation results shown in Figs.1e,f since MOISH provides a homodyne signal. The maximal orientation signal at t=T$_{rev}$ and not during, or in vicinity of the THz excitation is an intriguing signature for the resonant nature of the THz-dipole interaction, indicating that the molecules continue to accumulate rotational energy in a coherent manner throughout the entire interaction with the field and beyond the initial event of orientation around t=0. These rotational coherences manifest later on by enhanced orientation at t=T$_{rev}$, long after the THz field is over [8,17].

While the MOISH signals in Figs.1c,d are qualitatively in agreement with the theoretical predictions of Figs.1e,f (dashed red lines), in what follows we focus on their quantitative discrepancies. The main disagreement is revealed when comparing the ratio of the t=T$_{rev}$ signal ($S_{T_{rev}}$) and the t=0 signal ($S_0$) given by $R_{MOISH} = \int S_{T_{rev}} dt / \int S_0 dt$ to the simulated ratio $R_{Orient}^{theory} = \int |\tilde{O}_{T_{rev}}|^2 dt / \int |\tilde{O}_0|^2 dt$. The latter was found to be $R_{Orient}^{theory}$~2.8 and insensitive to the carrier-envelope phase (CEP, see SI.1) of the THz field whereas the experimental $R_{MOISH}$ in Figs.1c,d yields $R_{MOISH} = 1.8$. The discrepancy in Fig.1c of the relative peak intensities and the slight temporal shift between the experimental (black curve) and simulated results (dashed red curve) is readily observed and will become clearer in what follows.

In a set of measurements performed with different polar gas species at varying pressures we have found that $R_{MOISH}$ varies with both gas type and density of the gas, as shown in Fig.2 for three different gas species (CH$_3$I, OCS, N$_2$O) at the pressure range of 0-50torr. We note that figures 2a-c as well as Fig.4a share the same intensity scale (given in arb. units) demonstrating that the CH$_3$I signal is 1-2 orders of magnitude lager than those of OCS and N₂O. This large variation emanates from the first order ($\beta$) and second order ($\gamma$) hyperpolarizabilities of the different gasses as demonstrated by our calculations (for computational details see SI.3).

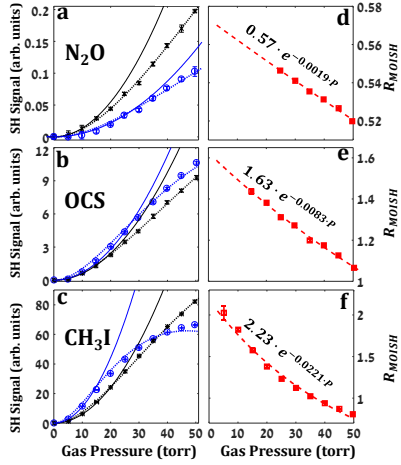

**Figure 2:** Experimental MOISH signals from different gasses at varying pressures. Figures a, b and c show the integrated signals at t=0 (black x's and dotted trend-line) and at t=1Trev (blue circles and dotted trend-line) from N₂O, OCS and CH₃I respectively. Figures d, e and f show the $R_{MOISH}$ of the signals in Figs.a-c, respectively. All measurements were performed with probe intensity of ~$3 \cdot 10^{13}$W/cm². The expected quadratic pressure dependence is depicted by the solid black and blue lines respectively.

Figures 2a-c depict the SHG measured at t=0 ($\int S_0 \mathrm{dt}$, black data points) and t=T$_{rev}$ ($\int S_{T_{rev}} \mathrm{dt}$, blue data points) for the three gas samples. The solid black and blue lines show the expected SHG dependence on pressure, produced by extrapolation of the quadratic fit of the first few (low pressure) data points. Trendlines of the experimental data sets are plotted by dotted curves. The deviation of the experimental SH from quadratic pressure dependence is attributed to:
(1) Collisional decoherence that effectively attenuates $S_{T_{rev}}$ and hardly affects $S_0$.
(2) Phase-mismatch experienced by the nonlinear SHG upon propagation in the gas.
Naturally, both these effects increase with gas density.

In order to ease the phase matching constraints we restrict the interaction length by placing an iris in the gas cell [32]. Furthermore, we note that the SHG signals at t=0 and t=T$_{rev}$ are affected similarly by phase-mismatch, thus their ratio ($R_{MOISH}$) is insensitive to phase-mismatch ramifications. Figures 2d-f depict the $R_{MOISH}$ obtained from the data in 2a-c, respectively. Owing to the collisional decoherence, $R_{MOISH}$ decays exponentially with pressure at a specific rate for each gas, in excellent agreement with those obtained via EOS (see SI.2) and reported in [17]. From the fitted exponential curves in Fig.2d-f we find that the collision-free $R_{MOISH}$ values (at $P = 0$, given by the pre-exponential factors) and $R_{Orient}^{theory} \sim 2.8$ remain in discrepancy ($R_{MOISH} = 0.57, 1.63, 2.2$ for N₂O, OCS, and CH₃I, respectively). This disagreement is attributed to the TFISH signal induced by the incident THz field (t=0) that constructively adds to the MOISH signal. The sum of these two contributions increases $S_0$ (at the denominator of $R_{MOISH}$) and results in lower $R_{MOISH}$ than that expected by orientation only. In what follows we analyze the nuclear (MOISH) and electronic (TFISH) contributions to the observed signals.

## **Selective contributions to the nonlinear susceptibility $\chi^{(2)}$**

A THz field force electronic oscillation in the field direction and transiently lifts the inversion symmetry of the medium, which provides an effective $\chi_{elect}^{(2)}$ detected via TFISH [35]. The latter, typically performed in air, is valid in polar molecules as well. In polar molecules however, the single-cycle THz induces further inversion asymmetry as it orients the molecules [7,8]. Thus, orientation yields an effective $\chi_{orient}^{(2)}$ medium that enables MOISH. Figure 3 depicts the expected electronic and nuclear contributions at t=0 and t=T$_{rev}$.

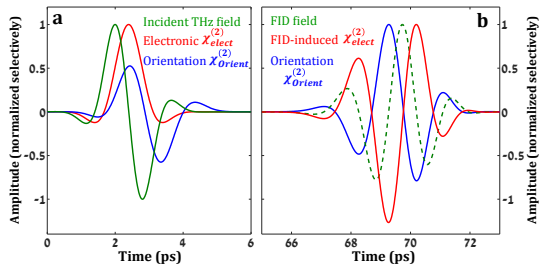

**Figure 3:** Simulation of the different $\chi^{(2)}$ contributions at (a) t=0 and (b) t=T$_{rev}$. The green curve depicts the incident THz field resulting in electronic (red curve) and the nuclear orientation (blue curve) contributions to the nonlinear susceptibility $\chi^{(2)}$.

The solid green curve in Fig.3a depicts the incident THz field as single-cycle pulse with anti-symmetric CEP, given by $E_{THz}(\mathrm{t}) = exp\left(-\frac{\mathrm{t}^2}{\sigma^2}\right) \cdot cos(\omega_0 \mathrm{t} + \pi/2)$. The parameters $\sigma = \frac{1.5\mathrm{ps}}{2\sqrt{\ln(2)}}$ and $\omega_0 = 0.5$THz were chosen to match our experimental THz field. The instantaneous electronic contribution induced by the field is calculated by the linear response $\chi_{elect}^{(2)}(t) \propto \chi^{(3)} \int_{-\infty}^{t} E_{THz}(t) \, dt$ and depicted by the (normalized) red solid line in Fig.3a. The solid blue curve depicts the nuclear contribution $\chi_{orient}^{(2)}$ upon orientation of the gas molecules $\langle \cos\theta \rangle_{(t)}$. The latter was simulated by numerically propagating the density matrix ,$\rho$, via the Liouville-Von Neumann equation $\frac{\partial \rho}{\partial t} = -\frac{i}{\hbar}[\hat{H}, \rho]$, with $\hat{H} = \frac{\hat{L}^2}{2I} + \hat{V}$ and $\hat{V} = -\vec{\mu} \cdot \vec{E}_{THz}(t)$ the dipole interaction term [8]. As shown in Fig.3a, the electronic and nuclear contributions at t=0 are coherent and 'in phase', thereby constructively adding to increase $\chi^{(2)}$ and enhance the SHG signal. We note that the incident THz field and electronic response (green and red respectively) are normalized in Fig.3a. The orientation response is normalized by the peak orientation at t=T$_{rev}$ (Fig.3b, blue curve). Fig.3b shows the different contributions to $\chi^{(2)}$ at t=T$_{rev}$. In addition to the orientation discussed above, we depict the electronic contribution of the emitted FID field. First, we derive the FID (dashed green curve in Fig.3b) from the transient orientation of the gas (blue curve in Fig.3b) by taking the time derivative of the latter [29,30]. Next, we calculate the linear response to the FID to obtain $\chi^{(2)}_{elect}$ (red curve in Fig.3b, normalized). The destructive interference of the two contributions is readily observed in Fig.3b (blue and red).

While the above analysis qualitatively explains the results of Fig.2, we note that in the discussion of Fig.2 we discarded the destructive electronic contribution induced by the FID at t=T$_{rev}$. Before we proceed to experimentally unveil the latter, let us consider the pressure-dependence of the different contributions discussed above:

At t=0 both the nuclear and electronic contributions are linear with the number of molecules in the interaction region: $\chi^{(2)}_{orient}(t=0) \propto \beta, P$ and $\chi^{(2)}_{elect}(t=0) \propto \gamma, P$. Here $\beta$ and $\gamma$ correspond to the electronic hyper-polarizabilities at the probe frequency (375THz) upon orientation and under the action of the THz respectively.

At *t=T$_{rev}$*, however, the two contributions differ in their pressure dependencies: while $\chi^{(2)}_{orient}(T_{rev}) \propto P$, the FID contribution depends on the density squared ($\propto P^2$) since it is induced by the emitted FID($\propto P$) that acts back on the same gas, hence $\chi^{(2)}_{elect}(T_{rev}) \propto P^2$. Note that the above refers to the SHG field ($E_{400nm}$) and not the (detected) intensity.

Furthermore, when comparing different gasses, we must consider their different dipole magnitudes; Consider a THz field $E_{THz}$ interacting through the molecular dipole, $\mu \cdot E_{THz}$. While the induced orientation $\langle cos\theta \rangle$ is linear with $\mu$, the FID that is emitted at t=T$_{rev}$ is quadratic with $\mu$ since $E_{FID} \propto -\frac{d\langle cos\theta \rangle}{dt} \cdot \mu \cdot P$ [18,40] where $\langle cos\theta \rangle \propto \mu$ (for experimental $E_{FID}$ vs. $\mu$ see SI.4). From all of the above we conclude that the extent to which the FID-induced $\chi^{(2)}_{elect}$ diminishes the MOISH signal at t=T$_{rev}$ (by destructively interfering with $\chi^{(2)}_{orient}$) depends on multiple factors: it increases with the hyperpolarizability ($\gamma$), the gas pressure ($\propto P^2$) and with the molecular dipole ($\propto \mu^2$).

Thus, for gasses with relatively low dipole such as N$_2$O ($\mu = 0.17D$) and OCS ($\mu = 0.7D$) the contribution of the FID is negligible and the decay rate of the signal at t=T$_{rev}$ is effectively governed by collisions as shown in Figs.2a,b. For larger dipole values (such as CH$_3$I with $\mu = 1.62D$) we expect to find a larger decay rate than that induced solely by collisions. Nevertheless, the decay of $R_{MOISH}$ in Fig.2c was found in good agreement with that quantified by EOS. This is attributed to the already large collisional decay of CH$_3$I that obscures the (relatively small) destructive contribution of the FID at t=T$_{rev}$. Thus, a large molecular dipole acts as a 'double-edged sword' since it increases the destructive FID contribution but enhances the collisional decay rate via dipole-dipole interactions and by that, obscures the FID contribution. Nevertheless, the first experimental indication for this elusive effect is presented in Figs.4a,b, where we conducted the exact same experiment of Fig.2 only for CH$_3$CN ($\mu = 3.92D$) and found that $R_{MOISH}$ decays ~30% faster than that quantified by EOS.

In order to experimentally validate the above hypothesis, one would like to vary the relative magnitudes of the two $\chi^{(2)}$ contributions selectively, however those are unavoidably inter-related. Instead, we utilized a non-polar gas that does not contribute to $\chi^{(2)}_{orient}$ but strongly affected by the FID of the polar gas, hence serves as a "reporter gas". This is done by injecting carbon-disulfide (CS$_2$) at varying partial densities in addition to the fixed density of the polar gas (CH$_3$I and CH$_3$CN in Figs.4c,d, respectively). For gas-mixing procedure see section SI.5.

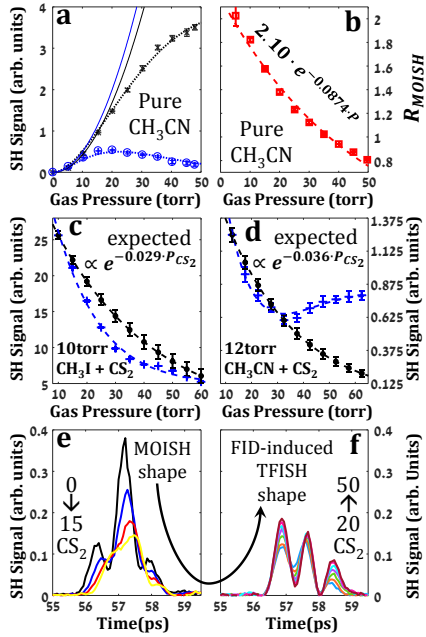

**Figure 4:** (a, b) same as in Fig.2 only for $CH_3CN$. (c) SH signals from a mixture of 10torr $CH_3I$ and varying $CS_2$ pressure. Calculated decay shown in black, experimental decay shown in blue. (d) same as (c) for 12torr $CH_3CN$ and varying $CS_2$ pressure. (e, f) time resolved signals of the data of figure (d) showing the evolution of the signal shape from that of MOISH (0→15torr $CS_2$) to that of FID-induced TFISH (20→50torr $CS_2$).

Figures 4c,d were obtained with fixed 10torr $CH_3I$, and 12torr $CH_3CN$ respectively, and varying $CS_2$ pressures. The black curves show the expected t=$T_{rev}$ signal of the mixture with collisional decay and phase matching effects accounted for (see section SI.6), but without the FID contribution. The blue data points (marked by '+') and dashed trend-lines depict the experimental results. In both gas mixtures, the FID emission interacts with the reporter $CS_2$ gas and induces $\chi^{(2)}_{elect}$ that partially counteracts the $\chi^{(2)}_{Orient}$ of the polar gas at t=$T_{rev}$ resulting in faster decay of the SH signal with increased $CS_2$ pressure. As the $CS_2$ pressure further increases, the decay rate of the SH gradually reduces and its trend reverses as the orientation- and FID- induced contributions become comparable (~30torr in Fig.4d). Above this pressure the SH signal starts to increase as the FID contribution overcomes that of the MOISH. This is shown in Figs. 4e,f where the shape of the time-resolved SHG signal gradually changes from that of MOISH (from $CH_3CN$ only) to that of the FID-induced TFISH (of both gasses). The incomplete destruction SH signal is due to the incomplete destructive interference of the two contributions emanating from the different rotational dynamics of the two gasses (evident from their time-resolved signals). In addition, the use of the reporter, non-polar gas provides yet another advantage over a pure polar gas sample: the decay of $R_{MOISH}$ is significantly reduced with $4 \cdot 10^{-3}$torr$^{-1}$ and $6 \cdot 10^{-3}$torr$^{-1}$ for the $CH_3I/CS2$ and $CH_3CN/CS2$ mixtures compared to $2.2 \cdot 10^{-2}$torr$^{-1}$ and $8.7 \cdot 10^{-2}$torr$^{-1}$ in neat gas, respectively (Figs.2c and 4b). The reduced decay rate improves the visibility of the FID contribution that is otherwise obscured by the rapid decay rate (due to strong dipole-dipole interactions in pure polar gas). We further note that in order to alleviate possible contributions of THz-induced rotational excitation of $CS_2$ owing to its large polarizability anisotropy [41,42], we repeated the reporter gas experiment with carbon-tetrachloride ($CCl_4$) and obtained very similar trends as in Figs.4c,d.

To conclude, we utilized the SH signal generated in THz-oriented gas phase molecules as a direct probe of orientation. The technique coined 'MOISH', is contributed by several electronic and nuclear (orientation) responses that together, govern the observed SH signal. These contributions were theoretically and experimentally explored in different gases and varying gas densities. A 'reporter gas' approach was developed to unveil the elusive (destructive) contribution of the secondary FID emission. MOISH offers a spatially localized, all-optical technique for direct probing of molecular orientation and provides new means for studying coherent rotational dynamics induced by concerted THz and optical excitations.

**Acknowledgements**
The authors thank Prof. Oded Hod and Prof. Michael Urbakh (TAU chemistry) for stimulating discussions. SF acknowledges the support of Israel Science Foundation (926/18) and the Wolfson Foundation (PR/ec/20419).

# Supplementary Information:
## Molecular Orientation-Induced Second Harmonic Generation: deciphering different contributions apart

## SI.1) The dependence of $R_{Orient}^{theory}$ on the CEP of the THz field

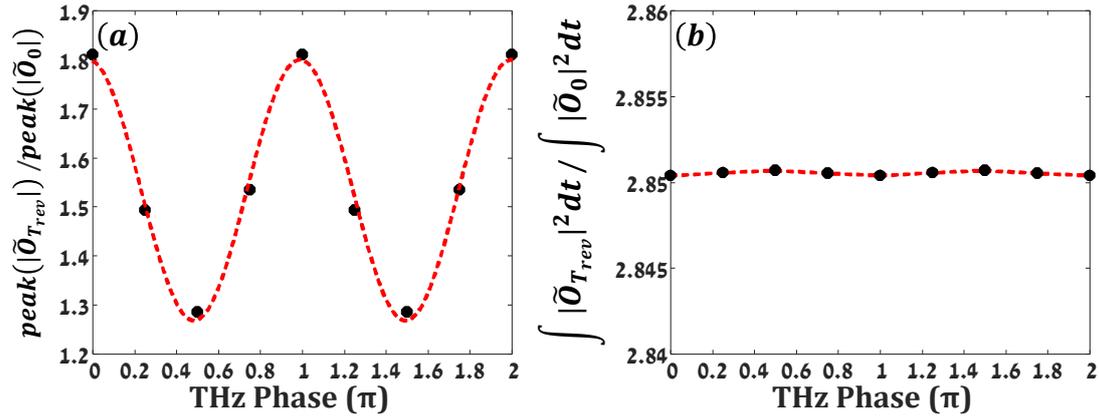

**Figure SI.1:** Simulated ratio between (a) peak orientation at t=1T$_{rev}$ and t=0 for CH$_3$I for varying CEP of the incident THz field demonstrating severe dependence on the latter. (b) Calculated $R_{Orient}^{theory} = \int |\tilde{O}_{T_{rev}}|^2 dt / \int |\tilde{O}_0|^2 dt$ remains fixed at 2.85 in all CEP range.

The temporal shape of orientation depends on the carrier-envelope phase (CEP) of the THz field [8,43]. While the $\langle cos\theta \rangle_{t=0}$ shape resembles that of the incident THz field, the shape of the $\langle cos\theta \rangle_{T_{rev}}$ transient is π/2 shifted with respect to the latter. Thus, the ratio between the peak orientation at $t = T_{rev}$ ($\tilde{O}_{T_{rev}}$) and at $t = 0$ ($\tilde{O}_0$) shows strong CEP dependence and may vary between ~1.3-1.8 with CEP variations (Fig.1a). However, by taking the ratio of the integrated orientation responses i.e. $\int |\tilde{O}_{T_{rev}}|^2 dt / \int |\tilde{O}_0|^2 dt$ one alleviates the CEP dependence and obtains a fixed ratio of ~2.85 (in CH$_3$I) [44]. Naturally the calculated ratio of 2.85 may slightly vary with the rotational coefficient ($B$), the temperature and spectral width of the excitation pulse.

## SI.2) Quantifying collisional decay rates via EOS measurements

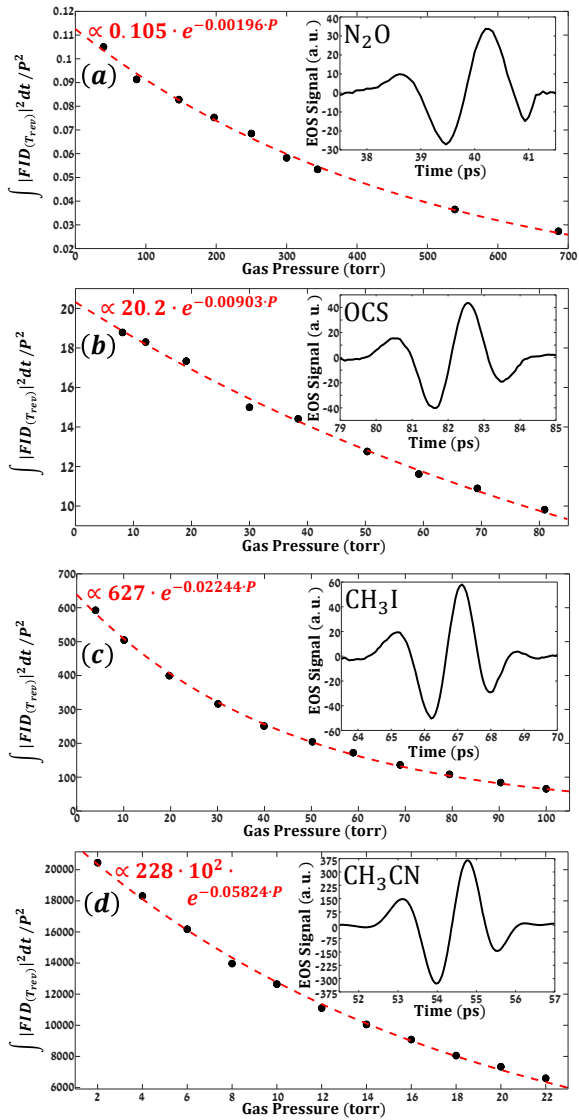

**Figure SI.2:** The pressure decay of the first FID signal (at $t = T_{rev}$) for different polar gases studied in this work: (a) $CH_3I$, (b) OCS, (c) $N_2O$ and (d) $CH_3CN$. The insets depict the typical time-resolved signal in each gas.

Figures 2d-f and Figure 4b in the main text file depict the pressure dependence of the experimental $R_{MOISH}$ that stands for the ratio of the integrated $t = 0$ and at $t = T_{rev}$ signals. Since the latter is subject to collisional decay and decoherence, we set to quantify the decay rate of the molecular orientation via THz-EOS measurement.

For each gas, we varied the pressure (P) and recorded the FID transients emitted at $t = T_{rev}$ in association with the orientation of the molecules.

The FID signal amplitude increases with the gas pressure P, and its exponential decay rate also increases linearly with P, i.e. $FID_{(T_{rev})} \propto Pe^{-\gamma \cdot P \cdot t}$ [18]. In order to extract the decay rates, $\gamma$ [sec$^{-1}$torr$^{-1}$], of the different gasses we integrated over each transient FID signal to obtain the signal area ($\int |FID_{(T_{rev})}|^2 dt$) and normalized by P2 to obtain the data points in Figs. SI.2a-d. We note that this quantification metric was shown resilient to centrifugal distortion effects [18,44]. The extracted exponential decay rates were found in excellent agreement with those obtained in [18] for $CH_3I, OCS, N_2O$ and with those obtained in $R_{MOISH}$ measurements of Fig.2a-c in the main paper file. We note that in CH3CN however, the decay rate of $R_{MOISH}$ (Fig.4b) is ~30% larger than in EOS (Fig SI.2d). This discrepancy in CH3CN serves as an indication for the destructive contribution of FID to the SH signal as discussed in details in the main text.

## SI.3) Theoretical calculations of Hyperpolarizabilities

The different contributions of $\chi^{(2)}_{elect}$ and $\chi^{(2)}_{orient}$ to the overall $\chi^{(2)}$ signal depends on the first (β) and second order (γ) hyperpolarizabilities respectively. These also dictate the clearly observed differences in signal amplitudes noted in Fig. 2.

The theoretical calculations of the hyperpolarizabilities were carried out using the density functional theory. In the calculations, the molecular geometric structures were optimized at the level of B3LYP/6-311G** [45,46]. For calculating the nonlinear optical (NLO) properties of the molecules of interest we used the hybrid GGA DFT methods reported to outperform pure DFT functionals [47,48] since the fraction of the Hartree-Fock (HF) exchange in the hybrid functionals can proportionally induce a modification on the asymptotic behavior of approximate exchange-correlation (XC) potentials and a reduction in the self-interaction error [49,50]. Accordingly, the hybrid functionals B3LYP (containing 20% of exact exchange) and PBE0 [51] (containing 25% of exact exchange) were chosen in this study for comparison. On the other hand, it was well known that large basis sets with polarization and diffuse functions played a significant role in a reliable estimation of the NLO properties. Herein, the diffuse function-augmented double(D)-ζ, triple(T)-ζ and quadruple(Q)-ζ basis sets aug-cc-pVnZ (n = D, T, and Q) from Dunning correlation-consistent basis set family [52–54], and the augmented triple-zeta and quadruple-zeta valence basis sets def2-nZVPD (n = T and Q) from Ahlrichs-Karlsruhe basis set family [55–58] were employed in the calculations for the first and the second hyperpolarizabilities based on their good performance on the linear and nonlinear properties of molecular materials [48,59,60].

For the first hyperpolarizability (β), the total molecular hyperpolarizability ($\beta_{tot}$) was defined as: $\beta_{tot} = \sqrt{\beta_x^2 + \beta_y^2 + \beta_z^2}$, where: $\beta_i = 1/3 \sum_j (\beta_{ijj} + \beta_{jji} + \beta_{jij})$, and where: $i,j = \{x, y, z\}$. Moreover, its projection on the dipole moment vector ($\beta_{prj}$), which was considered to be comparable to the data measured in the electric field induced second harmonic generation (EFISHG) experiment, was given by: $\beta_{prj} = \sum_i \frac{\mu_i \beta_i}{|\mu|}$. In terms of the second hyperpolarizability (γ), the total magnitude ($\gamma_{tot}$) was described as: $\gamma_{tot} = \sqrt{\gamma_x^2 + \gamma_y^2 + \gamma_z^2}$, where: $\gamma_i = 1/15 \sum_j (\gamma_{ijji} + \gamma_{ijij} + \gamma_{iijj})$. While the corresponding average magnitude ⟨γ⟩ was determined as: $\langle \gamma \rangle = \gamma_{||} = \gamma_x + \gamma_y + \gamma_z$.

All calculations were performed with Gaussian 16 package [61], while the NLO analysis was implemented by the means of Multiwfn 3.4.1 [62]. The calculated results for the first and the second hyperpolarizabilities are listed in Tables S1 and S2. According to the results, we can find that B3LYP and PBE0 perform similarly in the description of the NLO properties of molecules, while the results generally get converged at the level of basis sets with relatively large diffuse functions.

**Table S1.** The frequency-dependent first hyperpolarizability $\beta(-2\omega; \omega, \omega)$ ($\times 3.2063613061 \cdot 10^{-53} C^3 \cdot m^3 \cdot J^{-2}$) and the second hyperpolarizability $\gamma(-2\omega; \omega, \omega, 0)$ ($\times 6.2353799905 \cdot 10^{-65} C^4 \cdot m^4 \cdot J^{-3}$) of $CS_2$, $N_2O$, OCS, $CH_3I$, and $CH_3CN$ at wavelength of λ = 800nm calculated by B3LYP and PBE0 combined with different basis sets. For brevity, the basis sets aug-cc-pVDZ, aug-cc-pVDZ and aug-cc-pVDZ are marked as DZ, TZ, and QZ, while the prefix "def2" is omitted for the basis sets def2-TZVPD and def2-QZVPD. In addition, It should be noted that the first hyperpolarizabilities $\beta_{tot}$ and $\beta_{prj}$ of $CS_2$ are equal to 0.

| $CS_2$ | B3LYP | | | | | PBE0 | | | | |
|---|---|---|---|---|---|---|---|---|---|---|
| | λ = 800nm | | | | | λ = 800nm | | | | |
| | DZ | TZ | QZ | TZVPD | QZVPD | DZ | TZ | QZ | TZVPD | QZVPD |
| $\gamma_x$ | 2896.9 | 3997.28 | 5060.9 | 4166.94 | 4823.15 | 2481.31 | 3556.98 | 4424.85 | 3718.26 | 4209.02 |
| $\gamma_y$ | 6824.01 | 8384.31 | 9787.49 | 8627.42 | 9926.55 | 5841.68 | 7319.97 | 8446.47 | 7620.59 | 8604.59 |
| $\gamma_z$ | 2896.95 | 3997.35 | 5060.89 | 4166.87 | 2481.34 | 3557.08 | 4424.85 | 3718.22 | 4209 |
| $\gamma_{tot}$ | 7959.36 | 10112.1 | 12125.2 | 10447.9 | 12044.2 | 6814.63 | 8881.83 | 10512 | 9258.73 | 10462.8 |
| $\gamma_{||}$ | 12617.9 | 16378.9 | 19909.3 | 16961.2 | **19572.8** | 10804.3 | 14434 | 17296.2 | 15057.1 | 17022.6 |

| $N_2O$ | B3LYP | | | | | PBE0 | | | | |
|---|---|---|---|---|---|---|---|---|---|---|
| | λ = 800nm | | | | | λ = 800nm | | | | |
| | DZ | TZ | QZ | TZVPD | QZVPD | DZ | TZ | QZ | TZVPD | QZVPD |
| $\beta_{tot}$ | 72.38 | 64.708 | 60.696 | 44.986 | **52.178** | 70.01 | 62.982 | 59.662 | 46.182 | 52.274 |
| $\beta_{prj}$ | -72.38 | -64.708 | -60.696 | -44.986 | -52.178 | -70.01 | -62.982 | -59.662 | -46.182 | -52.274 |
| $\gamma_x$ | 344.416 | 428.796 | 520.361 | 309.923 | 390.231 | 311.712 | 395.387 | 474.556 | 293.393 | 362.682 |
| $\gamma_y$ | 344.416 | 428.796 | 520.361 | 309.923 | 390.231 | 311.712 | 395.387 | 474.556 | 293.393 | 362.682 |
| $\gamma_z$ | 856.978 | 964.236 | 1052.89 | 680.182 | 837.465 | 777.555 | 882.184 | 954.301 | 637.85 | 773.993 |
| $\gamma_{tot}$ | 985.726 | 1139.07 | 1284.57 | 809.168 | 1002.95 | 893.824 | 1044.47 | 1166.66 | 760.928 | 928.516 |

| γ∥ | 1545.81 | 1821.83 | 2093.61 | 1300.03 | **1617.93** | 1400.98 | 1672.96 | 1903.41 | 1224.64 | 1499.36 |
|---|---|---|---|---|---|---|---|---|---|---|
| **OCS** | \multicolumn{5}{c}{B3LYP} | | | | PBE0 | | |
| | \multicolumn{5}{c}{λ = 800nm} | | | | λ = 800nm | | |
| | DZ | TZ | QZ | TZVPD | QZVPD | DZ | TZ | QZ | TZVPD | QZVPD |
| $\beta_{tot}$ | 160.587 | 193.837 | 206.763 | 201.84 | 212.191 | 149.6 | 179.531 | 190.686 | 185.751 | 195.708 |
| $\beta_{prj}$ | -160.584 | -193.837 | -206.762 | -201.839 | -212.191 | -149.6 | -179.53 | -190.686 | -185.751 | -195.708 |
| $\gamma_x$ | 1500.49 | 2108.2 | 2709.82 | 2202.14 | 2556.46 | 1320.46 | 1912.17 | 2416.01 | 1994.74 | 2268.23 |
| $\gamma_y$ | 1498.91 | 2107.81 | 2710.62 | 2203.62 | 2558.02 | 1318.56 | 1911.99 | 2417.1 | 1995.9 | 2269.39 |
| $\gamma_z$ | 1533.34 | 2056.23 | 2451.71 | 2149.26 | 2465.45 | 1373.46 | 1852.33 | 2178.86 | 1931.32 | 2195.29 |
| $\gamma_{tot}$ | 2617.12 | 3621.53 | 4549.89 | 3784.8 | 4376.92 | 2317.03 | 3277.69 | 4053.01 | 3419.44 | 3887.71 |
| $\gamma_\parallel$ | 4532.74 | 6272.24 | 7872.15 | 6555.02 | **7579.93** | 4012.48 | 5676.48 | 7011.97 | 5921.96 | 6732.91 |
| **CH₃CN** | \multicolumn{5}{c}{B3LYP} | | | | PBE0 | | |
| | \multicolumn{5}{c}{λ = 800nm} | | | | λ = 800nm | | |
| | DZ | TZ | QZ | TZVPD | QZVPD | DZ | TZ | QZ | TZVPD | QZVPD |
| $\beta_{tot}$ | 14.233 | 25.911 | 30.97 | 6.996 | **20.742** | 15.545 | 24.115 | 27.929 | 7.793 | 19.401 |
| $\beta_{prj}$ | 14.233 | 25.911 | 30.97 | 6.996 | 20.742 | 15.545 | 24.115 | 27.929 | 7.793 | 19.401 |
| $\gamma_x$ | 2255.52 | 2481.31 | 2593.5 | 2030.16 | 2303.92 | 2075.32 | 2266.35 | 2357.55 | 1887.86 | 2120.81 |
| $\gamma_y$ | 1011.13 | 1235.81 | 1364.43 | 985.475 | 1204.03 | 945.471 | 1139.84 | 1248.39 | 929.193 | 1114 |
| $\gamma_z$ | 1011.22 | 1235.68 | 1363.55 | 985.433 | 1204.01 | 945.547 | 1139.95 | 1248.53 | 929.275 | 1114.13 |
| $\gamma_{tot}$ | 2670.64 | 3034.97 | 3232.21 | 2462.48 | 2864.85 | 2468.79 | 2781.2 | 2945.39 | 2300.21 | 2641.99 |
| $\gamma_\parallel$ | 4277.87 | 4952.8 | 5321.48 | 4001.07 | **4711.96** | 3966.33 | 4546.14 | 4854.47 | 3746.32 | 4348.95 |
| **CH₃I** | \multicolumn{5}{c}{B3LYP} | | | | PBE0 | | |
| | \multicolumn{5}{c}{λ = 800nm} | | | | λ = 800nm | | |
| | | | TZVPD | QZVPD | | | | TZVPD | QZVPD | |
| $\beta_{tot}$ | | | 361.347 | **376.31** | | | | 331.104 | 338.647 | |
| $\beta_{prj}$ | | | 361.347 | 376.31 | | | | 331.104 | 338.647 | |
| $\gamma_x$ | | | 6507.53 | 7452.54 | | | | 5917.21 | 6655.75 | |
| $\gamma_y$ | | | 6890.67 | 8588.62 | | | | 6254.16 | 7493.97 | |
| $\gamma_z$ | | | 6890.73 | 8589.32 | | | | 6255.76 | 7496.88 | |
| $\gamma_{tot}$ | | | 11718 | 14250.7 | | | | 10642.5 | 12516.5 | |
| $\gamma_\parallel$ | | | 20288.9 | **24630.5** | | | | 18427.1 | 21646.6 | |

Figure SI.3 depicts the experimental SHG signals obtained in Figs.2a-c and Fig.4.a ($\int S_0 dt$ and $\int S_{T_{rev}} dt$) extrapolated to $P = 0$, i.e. under decay and decoherence free conditions, for the four polar gasses vs. the calculated hyperpolarizabilities noted above. In order to compare the experimental results of the different gasses we must consider the degree of orientation induced by the THz field. The latter is linear with the permanent dipole of the molecules. Thus, the expected SHG signal contributed by the dipole orientation must be factored by the molecular dipole: $\chi^{(2)}_{orient-effective} \propto \mu \cdot \beta$ while the electronic contribution $\chi^{(2)}_{electronic} \propto \gamma$.

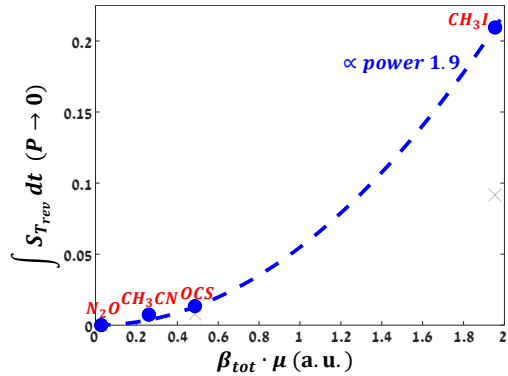

**Figure SI.3:** the figure depicts $\int S_{T_{rev}} dt$ extrapolated to $P = 0$ for the four polar gases plotted vs. $\chi^{(2)}_{orient-effective}$.

Figure SI.3 show that $\int S_{T_{rev}} dt \sim \left(\chi^{(2)}_{orient-effective}\right)^{1.9}$ in agreement with the expected quadratic dependence. At t=0 (not shown here) the signal includes both the nuclear and electronic contributions, that cannot be separated and therefore cannot be fitted selectively. These SH signal magnitude of the four polar gases (Figs.2a-c and Fig.4.a in the main text): $SH_{(CH_3I)} \gg SH_{(OCS)} > SH_{(N_2O)} > SH_{(CH_3CN)}$, is in good correspondence with the calculated β.

**SI.4) FID dependence on the molecular dipole**

In what follows we provide experimental verification for the quadratic dependence of the FID field on the molecular dipole ($E_{FID} \propto \mu^2$) using the pressure dependent FID transients measured via THz-EOS (SI.2). Consider the pre-exponential factors of the fits in Fig. SI.2. In fact, these pre-factors represent the integrated intensity per torr² $\int |E_{FID}|^2 dt/P^2$ of the FID, extrapolated to $P=0$, i.e. under decay and decoherence free conditions. Since all four gas samples were measured under the exact same experimental conditions (Incident THz amplitude, temperature, interaction volume), their ratio corresponds to the ratio of the emitted energies. Thus, by plotting the extracted pre-exponential factors of Fig.SI2 vs. the dipole and fitting to a power law, one obtains $\propto \mu^{4.07}$ power dependence which corresponds to quadratic dependence of $E_{FID}$ on the molecular dipole ($E_{FID} \propto \mu^2$).

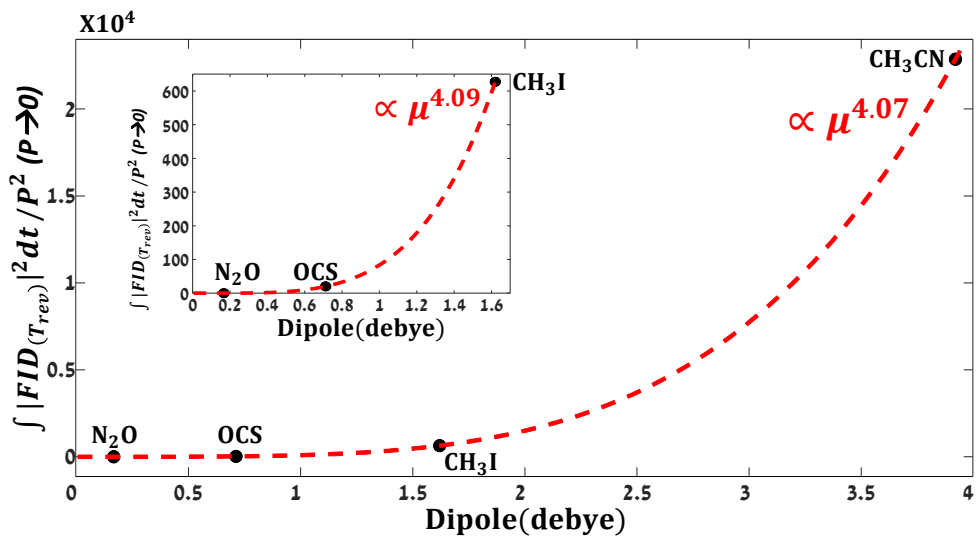

**Figure SI.4:** Extrapolated collision-free FID energy vs. the molecular dipole, showing $\propto \mu^4$ power law dependence.

To further verify the quartic power dependence that relies on 4 data points (4 gas types studied), we note (and show in the inset) that the fit results for the first three gasses yielded the same power law dependence.

**SI.5) Gas mixing procedure**

Gas mixtures with predetermined partial pressures of polar ($CH_3I$ or $CH_3CN$) and non-polar ($CS_2$) "reporter gas" were prepared via the following procedure using the experimental system shown in Fig. SI5. The setup includes a gas manifold connected to our static gas cell modified to include a 'cold finger'.

First, the cell was evacuated by the vacuum pump. Then, a fixed pressure of the polar gas (10torr for $CH_3I$ or 12torr for $CH_3CN$) was introduced into the entire manifold via the sample input port equipped with a Wilson seal [63]. Next, the cell valve was closed and the 'cold finger' cooled down by liquid nitrogen to condense the gas at the 'cold finger'. The manifold was evacuated down to <<0.1 torr. A small amount of $CS_2$ was introduced to the manifold and the manifold pressure recorded. Using our pre-calibrated ratio of the manifold volume and the cell volume, we can determine the exact pressure required in the manifold in order to achieve the desired pressure in the gas cell. By opening the cell valve we let the reporter gas condensate at the 'cold finger' until the pressure gauge reads '0' (namely <<0.1torr in our case). We close the cell valve and remove the liquid nitrogen from the cold finger, allowing it to thermalize and the gasses to homogeneously evaporate in the closed gas cell.

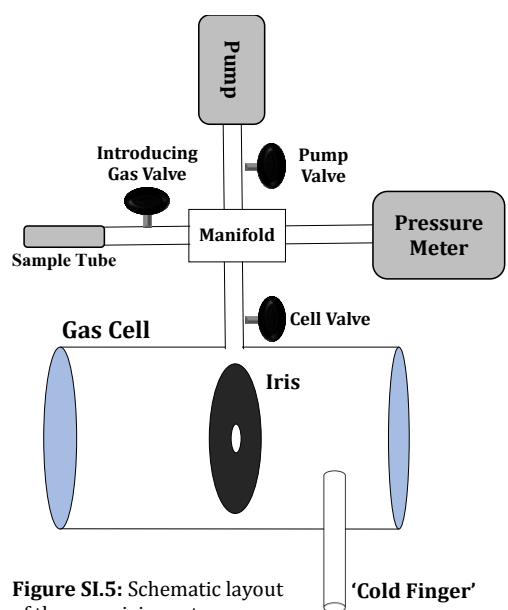

**Figure SI.5:** Schematic layout of the gas mixing setup.

**SI.6) Phase-mismatch and collisional decay of MOISH in gas mixtures**

Figures 4c,d in the main text compare the expected MOISH signals to the experimentally measured signals in the gas mixtures. In what follows we describe how we produced the expected curve (dashed black and data points).

The expected curve uses the SH signal measured in the pure polar gas at the fixed pressure used in the mixture, namely 10torr for $CH_3I$ and 12torr for $CH_3CN$. These 'first' data points already suffer from collisional decay and phase mismatch of the pure polar gas. Thus, our task is to quantify the ramifications of the added $CS_2$ imparted on the SH signal within the mixture. These are the collisional decay rate $\gamma_{CS_2}$ and the phase mismatch $\Delta k_{CS_2}$.

**a) Quantifying collisional decay rate of the bimolecular mixture:**
The collisional decay rate induced by $CS_2$ to the rotational dynamics of the polar gas was measured via EOS in the binary gas mixtures of interest ($CH_3I$ and $CS_2$, $CH_3CN$ and $CS_2$). The FID field emanates solely from the polar gas which is kept at fixed density. Therefore, by monitoring the FID field under varying $CS_2$ pressures, one directly obtains the $CS_2$-induced decay rate of interest.

Figures SI.6a,b show the extracted $CS_2$-induced decay of the orientation (FID) of $CH_3I$ ($\gamma_{CS_2} = 4.075 \cdot 10^{-3}$ torr$^{-1}$ and $CH_3CN$ ($\gamma_{CS_2} = 5.74 \cdot 10^{-3}$ torr$^{-1}$) respectively. We note that as in section SI.2, also here we integrate over the square of the FID field and fit to exponential decay as noted in the figures.

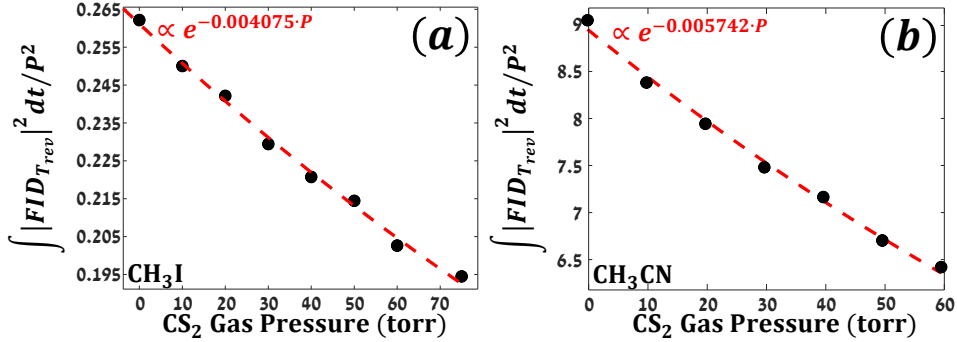

**Figure SI.6ab:** Figures a and b depict the EOS measurements done in gas-mixture of 10torr $CH_3I$/12torr $CH_3CN$ and with varying $CS_2$ pressures respectively.

An important advantage of the 'reporter gas' approach is the reduced decay rate imposed by $CS_2$ on the polar gas dynamics compared to the decay rate of pure polar gasses: while the pure polar gasses decay with $\gamma^{CH_3I} \sim 0.022$torr$^{-1}$ and $\gamma^{CH_3CN} \sim 0.058$torr$^{-1}$ due to strong dipole-dipole interactions, they decay with $\gamma_{CS_2}^{CH_3I} \sim 0.004$torr$^{-1}$ and $\gamma_{CS_2}^{CH_3CN} \sim 0.006$torr$^{-1}$. The drastically reduced decay rate is yet another advantage that increases the visibility of the FID contribution to the SH signal via the "reporter-gas" technique.

**b) Quantifying the $CS_2$-induced phase mismatch ($\Delta k$)**
The expected curves of Fig.4c,d in the main text use the experimentally measured signal of the pure polar gas as the basis of the SH signal at $t = T_{rev}$ and account for the prospected decay of the signal due to collisions with the added $CS_2$ (SI.6a,b) and the phase mismatch it imposes as described in this section. In fact, the phase mismatch introduced by 10torr of $CH_3I$ or 12torr $CH_3CN$ are inherently included in the first data point of each curve, thus our mission reduces to extract the $\Delta k$ contributed by $CS_2$ at the varying pressures. For this task we performed a set of TFISH measurements in neat $CS_2$ gas at varying pressures.

The SH signal from the nonpolar $CS_2$ gas depends on the gas pressure: $E_{SH} \propto P_{CS_2} \cdot e^{-\Delta k_{CS_2} \cdot L \cdot P_{CS_2}}$
Namely, the emitted SH field is proportional to the gas pressure. The phase-mismatch $e^{-\Delta k_{CS_2} \cdot L \cdot P_{CS_2}}$ also increases with pressure (here $\Delta k_{CS_2}$ is given in [cm$^{-1} \cdot$ torr$^{-1}$] and L is the effective length of interaction). By fitting our experimental results to a set of TFISH measurements at varying $CS_2$ pressures we can extract $\Delta k_{CS_2}$. Specifically, since we measure the SH intensity $|E_{SH}|^2$, we fit our data points to $I_{SH} \propto P_{CS_2}^2 \cdot e^{-2\cdot\Delta k_{CS_2}\cdot L\cdot P_{CS_2}}$ as shown in Figs. SI6c,d.

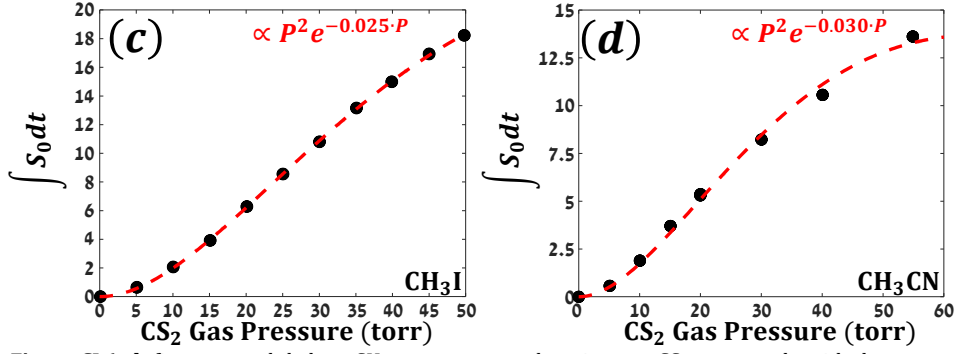

**Figure SI.6cd:** figures c and d show SH measurements done in pure $CS_2$ gas sample with the same experimental configurations of their corresponding figures 4c,d in the main paper file.

Figures SI.6cd shows the integrated TFISH signals of pure $CS_2$ gas at varying pressures. It may seem peculiar that these two, allegedly similar, data sets provide slightly different phase-mismatch parameters. However, we remind the readers that our experimental gas cell is equipped with an iris that serves to ease the phase matching constraints as reported previously [32]. This iris effectively reduces the interaction length by blocking the THz beam around the focus of the optical probe. The exact position of the iris is determined by the optimal (maximal) MOISH signal of the pure polar sample ($CH_3I$ and $CH_3CN$ in c,d respectively). Once the iris is set in position, we evacuate the cell and perform the $CS_2$ TFISH measurements, keeping the iris position fixed throughout all measurements of the pure and mixed samples. Since the exact iris position may be slightly differ for each sample, the retrieved phase mismatch of the $CS_2$ may slightly differ with different samples as shown by the experimental results.

With both the decay rate of the mixtures extracted from EOS measurements (SI.6a,b) and the phase mismatch (for specific iris configuration) extracted from TFISH measurements of the $CS_2$ reporter gas in hand, we can calculate and plot the expected SH signal as noted in the following table:

| Gas sample | Measured SH $\int S_{T_{rev}}dt$ (a.u.) | Fitted phase-mismatch $2\Delta k_{CS_2} \cdot L$ (torr$^{-1}$) | Collision decay $2\gamma_{mixture} \cdot T_{rev}$ (torr$^{-1}$) | Expected signal |
|---|---|---|---|---|
| $CH_3I$ (10 torr) | 25.6 | 0.02492 | 0.004075 | $\mathbf{25.6 \cdot e^{-0.029\cdot P_{CS_2}}}$ |
| $CH_3CN$ (12 torr) | 1.25 | 0.03044 | 0.005742 | $\mathbf{1.25 \cdot e^{-0.036\cdot P_{CS_2}}}$ |
| **Table 1:** Phase-matching and collisional decay rates extracted from Figs.SI6. | | | | |

15